\documentclass[10pt,a4paper,fleqn]{scrartcl}

\usepackage{amsmath}

\usepackage{amssymb}

\usepackage{color}

\usepackage{graphicx}

\usepackage[compress,authoryear]{natbib}

\usepackage{fancyhdr}

\usepackage{url}

\title{R\&D Strategy Document}
\subtitle{A Paper by the Olsen Ltd. Research Group}
\author{James B. Glattfelder, Thomas Bisig \\ and Richard B. Olsen}
\date{Olsen Ltd.\\ Seefeldstrasse 233, CH-8008 Z\"{u}rich
\\ \ \\ December 16, 2010 }%$3^{rd}$ of March 2010}

\begin{document}

\maketitle

\thispagestyle{empty}
\pagestyle{empty}

\clearpage
\pagebreak

%\pagenumbering{roman}

\begin{abstract}
\begin{it}
{\bf Research:} the process of discovering fundamental new knowledge and understanding.

{\bf Development}: the process by which new knowledge and understanding is applied.

\end{it}
\end{abstract}

\tableofcontents

\clearpage

\setcounter{page}{1}

\pagestyle{fancy}
\renewcommand{\subsectionmark}[1]{}   %rwmove subsection on rheader
\renewcommand{\sectionmark}[1]{\markboth{\thesection.\ #1}{}}   %not in caps
\rhead{\thepage}
\lfoot{\footnotesize \it Olsen Ltd.}
\rfoot{\footnotesize \it December 16, 2010} %\today} %March 2010}
\cfoot{}

%%%

%\pagenumbering{arabic}

%%%
\section{Introduction}

We provide an overview of the R\&D focus at the Zurich cell of Olsen
Ltd. By detailing the general conceptual framework and by identifying
key themes embedded in it, we outline what we believe are the
prerequisites and building-blocks for successfully devising trading
models (TMs) and other financial applications.

Chapters \ref{sec:lon}, \ref{sec:fund}, and \ref{sec:cs} are based on Appendix A,
Chapter \ref{sec:sl} on Appendix C of \cite{glattfelder10thesis}.

This document is the property of Olsen Ltd. and is proprietary and confidential. It is not
permitted to disclose, copy or distribute the document except by written permission of Olsen Ltd.

%%%
\section{In a Nutshell: Science and Laws of Nature}
\label{sec:lon}

Laws of nature can be seen as regularities and
structures in a highly complex universe. They depend critically on
only a small set of conditions, and are independent of many other
conditions which could also possibly have an effect.

Science can be understood as a quest to capture laws of nature within
the framework of a formal representation, or model. Naively one would
expect science to adhere to basic notions of common sense, like logic,
empiricism, causality, and rationality.

The philosophy of science deals with the assumptions,
foundations, methods and implications of science. It tries to describe
what constitutes a law of nature and tries to answer the question of
what knowledge is and how it is acquired.

In the philosophy of science, the programs of logical empiricism and
critical rationalism have been unsuccessful in conclusively answering
the above mentioned questions and in providing an ultimate
justification for science based on common sense. Indeed, the streams
of postmodernism, constructivism and relativism explicitly question
the notions of objectivity, rationality, absolute truths and
empiricism.
 
But then, why has science been so successful at describing reality?
And why is science producing the most amazing technology at breakneck
speed? It is a great feature of reality, that the formal models which
the human mind discovers/devises find their match in the workings of
nature. We will return to this enigma later on.

In being pragmatic and disregarding the conceptual problems, one can
identify two domains of nature and two modes of describing reality. In
the following section, we give a short overview of this {\it
  weltanschauung}. In Secs. \ref{sec:fund} and \ref{sec:cs} the
details are described.

\subsection{Overview: The Outline at a Glance}

The functioning of nature can broadly speaking be separated into two
categories, either belonging to the domain of {\it fundamental
  processes} or {\it complex systems}.  As examples, elementary
particles in a force-field represent the former, while a swarm of
birds constitutes the latter.

It has been possible to describe nature with two methods: the {\it
  analytical} and {\it algorithmic} approach. The analytical approach
is what most people are familiar with. Physical problems are
translated into mathematical equations which, when solved, give new
insights. The algorithmic approach simulates the physical system in a
computer according to algorithms, where the dynamics of the real
system are described by the evolution of the simulation.

\begin{figure}[tH]
\centering
 \includegraphics[width=0.5\textwidth,angle=0]{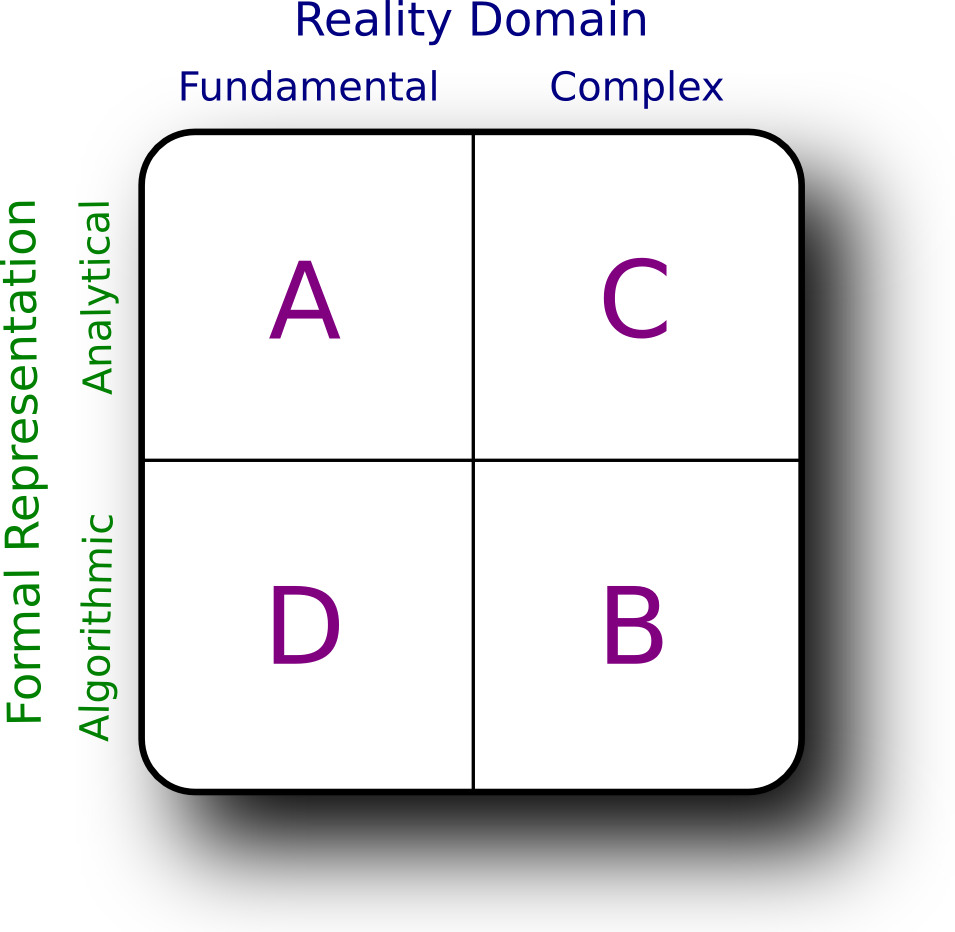}
\caption{Overview of the understanding of the laws of nature: domain of reality vs. possible formal representations. Taken from \cite{glattfelder10thesis}.} 
\label{tbl:ovwerlaws}
\end{figure}

In Fig. \ref{tbl:ovwerlaws} a simple illustration of this
categorization is seen. Four combinations emerge: both fundamental
processes and complex systems can be tackled analytically or
algorithmically. 

An obvious challenge is to identify the most successful method to
investigate a certain problem. This means not only choosing the formal
representation but also identifying the reality domain it belongs to.
Most of science can be seen to be related to strategy {\bf \textsf A},
as is discussed in Sec. \ref{sec:fund}.  Strategy {\bf \textsf B} has
been successful in addressing real-world complex systems, which is
detailed in Sec. \ref{sec:cs}.

The possibilities {\bf \textsf C} and {\bf \textsf D} have only been
sparsely explored. Regarding the former, some authors have recently
argued that complex systems can and should be tackled with
mathematical analysis. See for instance \cite{sornette2008ise}.
Strategy {\bf \textsf D} is mostly uncharted, and some tentative
efforts include describing space-time as a network in some fundamental
theories of quantum gravity (e.g., spin networks in loop quantum
gravity) or deriving fundamental laws from cellular automaton networks
\cite{NKS}.

We make two choices we believe to be instrumental to the success of
understanding real-world financial markets and devising profitable TMs:
\begin{enumerate}
\item understand the problem as originating
 from the domain of complex systems;
\item tackle the problem with an algorithmic approach.
\end{enumerate}
Hence, from our point of view, we see {\bf \textsf B} as the most
promising strategy, see also Sec. \ref{sec:fx}.

%%%
\section{Fundamental Processes and Mathematical Models ({\textsf A})}
\label{sec:fund}

As mentioned, science can be understood as the quest to capture the
processes of nature within formal mathematical representations.  In
other words, ``mathematics is the blueprint of reality'' in the sense
that formal systems are the foundation of science.

\subsection{The Success of Mathematical Models}

\begin{figure}[tH]
       \begin{center}
                \includegraphics[scale=0.45]{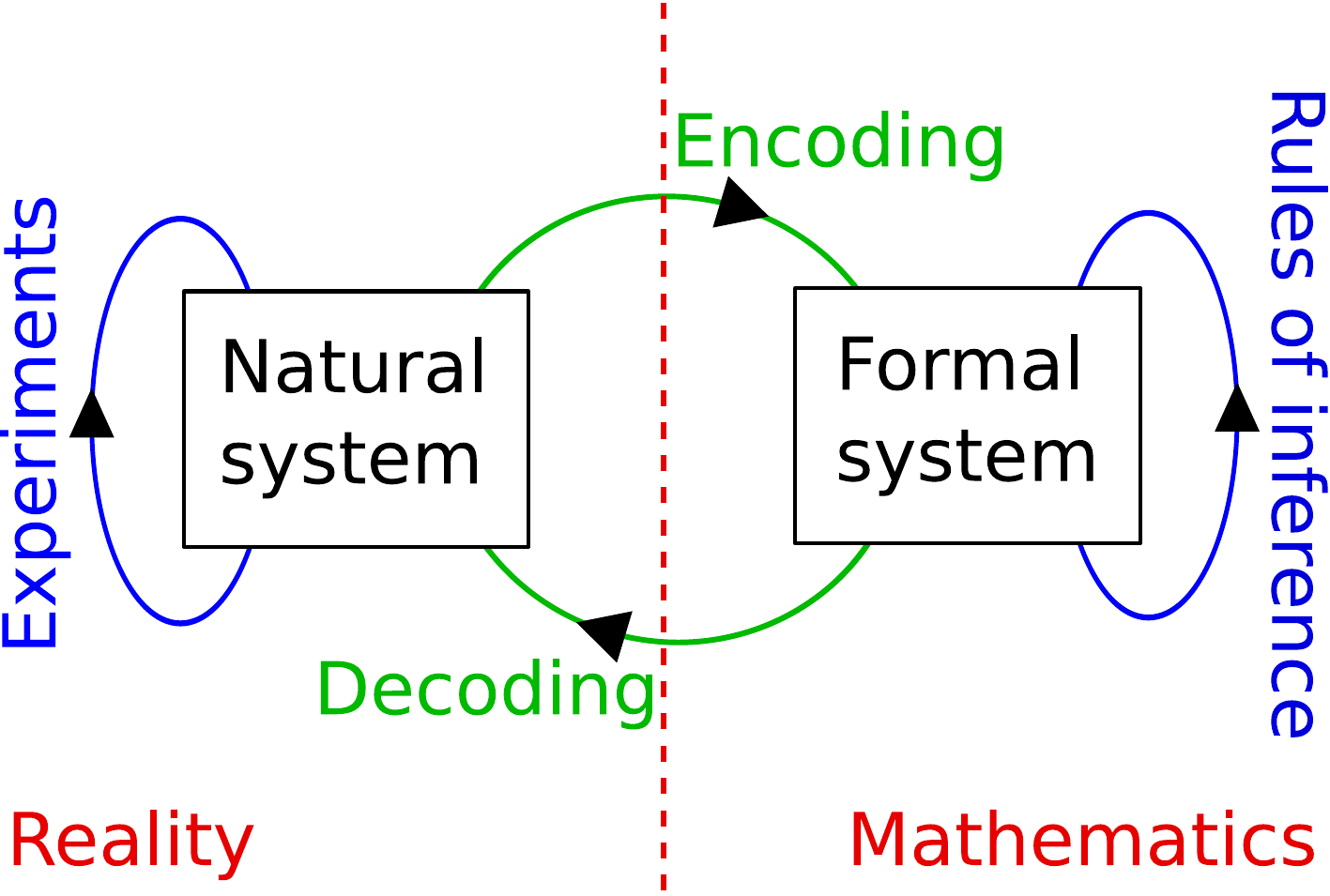}
                \caption{Schematic illustration of the interplay
                  between the outer reality and the mind's reality
                  harboring formal thought systems, following
\cite{casti1989arm}; see main text for a discussion. Taken from \cite{glattfelder10thesis}.}\label{fig:fund}
        \end{center} 
\end{figure}

This notion is illustrated in Fig. \ref{fig:fund}. The left-hand side
of the diagram represents the real world, i.e., the observable
universe.  Scientists focus on a well-defined problem or question,
identifying a relevant subset of reality, also called a natural
system. To understand more about the nature of the natural system
under investigation, experiments are performed yielding new
information.  Robert Boyle was instrumental in establishing
experiments as the cornerstone of physical sciences around 1660.
Approximately at the same time, the philosopher Francis Bacon
introduced modifications to Aristotle's (nearly two thousand year old)
ideas, introducing the so-called scientific method where inductive
reasoning plays an important role. This paved the way for a modern
understanding of scientific inquiry.

In essence, guided by thought, observation and measurement, natural
systems can be ``encoded'' into formal systems, depicted on the
right-hand side of Fig. \ref{fig:fund}. Representing nature as
mathematical abstractions is understood as a mapping from the real
world to the mathematical world. Using logic (e.g., rules of
inference) in the formal system, predictions about the natural system
can be made. These predictions can be understood as a mapping back to
the physical world, ``decoding'' the knowledge gained from from the
abstract model. Checking the predictions with the experimental outcome
shows the validity of the formal system as a model for the natural
system.

The following two examples should underline the great power of this
approach, where new features of reality where discovered solely on the
requirements of the mathematical model. Firstly, in order to unify
electromagnetism with the weak force (two of the three
non-gravitational forces), the theory postulated two new elementary
particles: the W and Z bosons. Needless to say, these particles where
hitherto unknown and it took 10 years for technology to advance
sufficiently to prove their existence.  Secondly, the fusion of
quantum mechanics and special relativity lead to the Dirac equation
which demands the existence of an, up to then, unknown flavor of
matter: antimatter. Four years after the formulation of the theory,
antimatter was experimentally discovered.

\subsection{The Paradigm of Fundamental Processes}

Is it possible to isolate a single idea that has been instrumental to
the success of describing fundamental processes? 

It can be argued that the most fruitful paradigm in the study of
fundamental processes in nature has been:
\begin{quote} \bf
    P1. Mathematical models of reality are independent of their formal representation.
\end{quote}
This idea leads to the notions of symmetry and invariance.

To illustrate, imagine an arrow located in space. It has a length and
an orientation. In the mathematical world, this can be represented by
a vector, labeled $a$. By choosing a coordinate system, the abstract
entity $a$ can be given physical meaning, for instance $a = (3, 5,
1)$.  The problem is however, that depending on the choice of the
coordinate system, which is arbitrary, the same vector is described
very differently: $a = (3, 5, 1) = (0, 23.34, -17)$. The paradigm
above states that the physical content of the mathematical model
should be independent form the choice of how one represents the
mathematical model.

Although this sounds rather trivial, for physics it has very deep
consequences. For instance, the requirements that physical experiments
should be unaffected by the time of day and geographic location they
are performed at, are formalized as time and translational invariance,
respectively. These requirements alone give rise to the conservation
of energy and momentum (Noether's theorem).

\begin{figure}[tH]
       \begin{center}
                \includegraphics[scale=0.4]{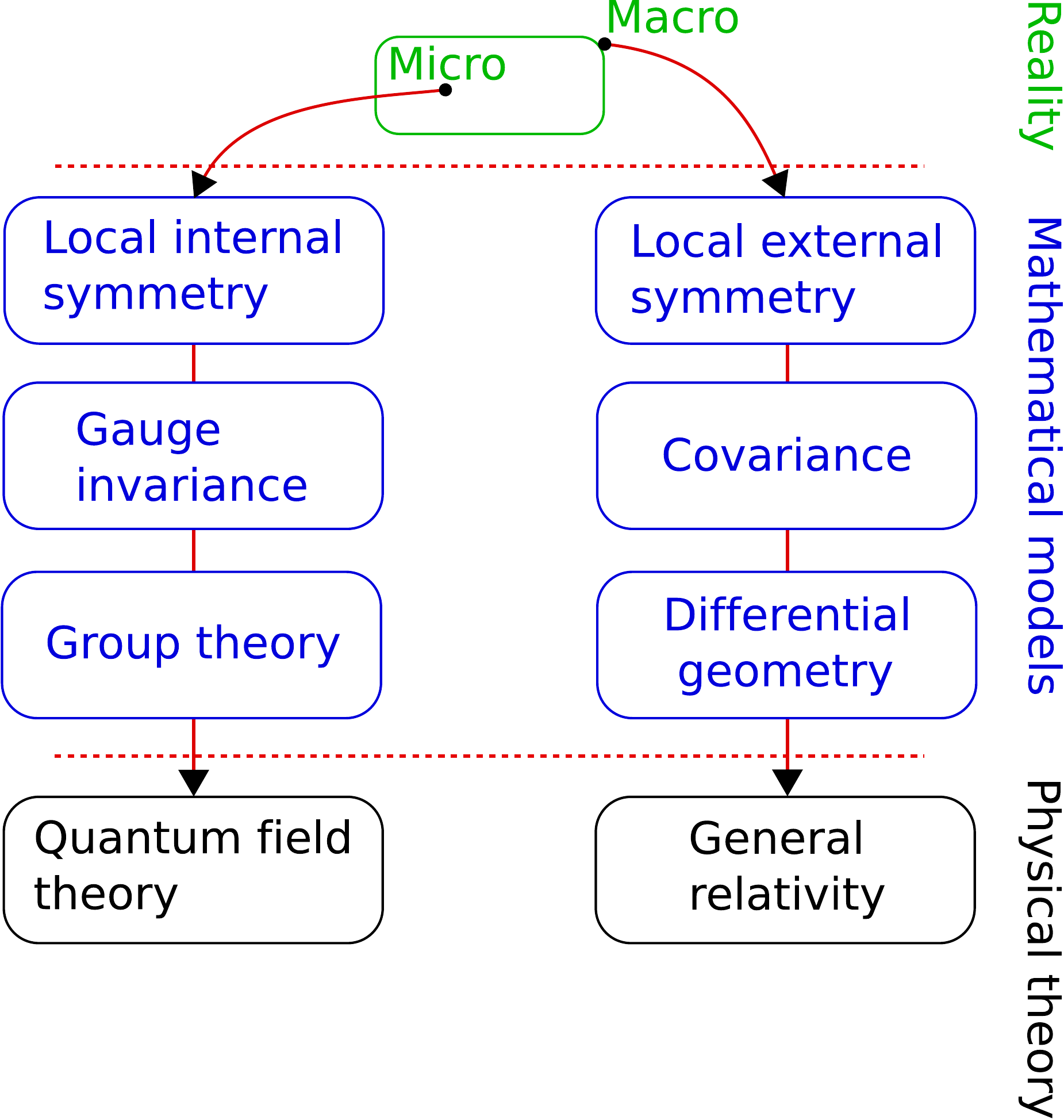}
                \caption{A big chunk of reality described by the symmetry of
mathematical models. Taken from \cite{glattfelder10thesis}.}\label{fig:mm}
        \end{center}
\end{figure}

Generally, the requirement of invariance and symmetry results in a
great part of physics, from quantum field theories to general relativity,
see Fig.  \ref{fig:mm}.

In general relativity the vectors are somewhat like multidimensional
equivalents called tensors and the common sense requirement, that the
calculations involving tensor do not depend on how they are
represented in space-time, is covariance. See the right-hand side of
Fig. \ref{fig:mm}. It is quite striking, but there is only one more
ingredient needed in order to construct one of the most aesthetic and
accurate theories in physics. It is called the equivalence principle
and states that the gravitational force is equivalent to the forces
experienced during acceleration. Again, this may sound trivial, has
however very deep implications.

The Standard Model of elementary particle physics unites the quantum
field theories describing the fundamental interactions of particles in
terms of their (gauge) symmetries. See the left-hand side of Fig.
\ref{fig:mm}.

\subsection{The Challenges}

We now come back to the questions raised in Sec. \ref{sec:lon}.  Why
has science been so successful in describing reality and why is
science producing amazing technology at breakneck
speed?\footnote{Especially when recalling that the attempts to found
  science on common sense notions have been unsuccessful.}

The simple answer is that there is no reason for this to be the case,
other than the fact that it is the way things are. The Nobel laureate
Eugene Wigner captures this salient fact in his essay ``The
Unreasonable Effectiveness of Mathematics in the Natural Sciences'':
\cite{wigner1960unreasonable}:
\begin{quote} \it
   `` [\dots] the enormous usefulness of mathematics in the natural sciences is something bordering on the mysterious and [\dots] there is no rational explanation for it.''

     ``[\dots] it is not at all natural that `laws of nature' exist, much less that man is able to discover them.''

     ``[\dots] the two miracles of the existence of laws of nature and of the human mind's capacity to divine them.''

     ``[\dots] fundamentally, we do not know why our theories work so well.''
\end{quote}

Over the past 300 years, physics has been very successful with this
approach, describing most of the observable universe. In essence, this
formalism works well for the fundamental workings of nature (strategy
{\bf \textsf A}).

However, to explain real-life complex phenomena, one needs to adopt a more
systems oriented focus. This also means that the interactions of
entities becomes an integral part of the formalism.

Some ideas should illustrate the change in perspective:
\begin{itemize}
\item most calculations in physics are idealizations and neglect dissipative effects like friction;
\item most calculations in physics deal with linear effects, as non-linearity is hard to tackle and is associated with chaos; however, most physical systems in nature are inherently non-linear \cite{strogatz1994non};
\item the analytical solution of three gravitating bodies in classical mechanics, given their initial positions, masses, and velocities, cannot be found; it turns out to be a chaotic system which can only be simulated in a computer; there are an estimated hundred billion galaxies in the universe.
\end{itemize}

%%%
\section{Complex Systems and the Algorithmic Approach ({\textsf B})}
\label{sec:cs}

A {\it complex system} is usually understood as being comprised of
many interacting or interconnected parts. A characteristic feature of
such systems is that the whole often exhibits properties not obvious
from the properties of the individual parts. This is called {\it
  emergence}. In other words, a key issue is how the macro behavior
emerges from the interactions of the system's elements at the micro
level. Moreover, complex systems also exhibit a high level of
adaptability and self-organization. The domains complex systems
originate from are mostly socio-economical, biological or
physio-chemical.

The study of complex systems appears complicated, as it is different
to the reductionistic approach of established science.  A quote from
\cite{anderson1972more} illustrates this fact:
\begin{quote}
\it
``At each stage [of complexity] entirely new laws, concepts, and
generalizations are necessary [\dots]. Psychology is not applied
biology, nor is biology applied chemistry''.
\end{quote}
This means that the knowledge about the constituents of a system
doesn't reveal any insights into how the system will behave as a
whole; so it is not at all clear how you get from quarks and leptons
via DNA to a human brain and consciousness.

Moreover, complex systems are usually very reluctant to be cast into
closed-form analytical expressions. This means that it is generally
hard to derive mathematical quantities describing the properties and
dynamics of the system under study.

%%%
\subsection{The Paradigms of Complex Systems}

The paradigms of complex systems are straightforward:
\begin{quote}
\bf
PI. Every complex system is reduced to a set of objects and a set of
functions between the objects.

PII. Macroscopic complexity is the result of simple rules of interaction at
the micro level.
\end{quote}

Paradigm I is reminiscent of the natural problem solving philosophy of
object-oriented programming, where the objects are implementations of
classes (collections of properties and functions) interacting via
functions (public methods). A programming problem is analyzed in terms
of objects and the nature of communication between them. When a
program is executed, objects interact with each other by sending
messages. The whole system obeys certain rules (encapsulation,
inheritance, polymorphism, etc.).

Indeed, in mathematics the field of category theory defines a category
as the most basic structure: a set of objects and a set of morphisms
(maps between the sets) \cite{hillman2001categorical}.  Special types
of mappings, called functors, map categories into each other. Category
theory was understood as the ``unification of mathematics'' in the
1940s.

A natural incarnation of a category is given by a complex network
where the nodes represent the objects and the links describe their
relationship or interaction. Now the structure of the network (i.e.,
the topology) determines the function of the network.  There are many
excellent introductory texts, surveys, overviews and books covering
the many topics related to complex networks:
\cite{strogatz2001ecn,albert2002smc,dorogovtsev2002evolution,DM03,newman2003structure,newman2006structure,caldarelli2007sfn,costa2007characterization,vega2007complex}.

This also highlights the paradigm shift from mathematical (analytical)
models to algorithmic models (computations and simulations performed
in computers).  In other words, the analytical description of complex
systems is abandoned in favor of algorithms describing the interaction
of the objects, also called {\it agents}, in a system according to
rules of local interactions. This approach has given rise to the
prominent field of {\it agent-based modeling}. In addition, a key
realization is that also the structure and complexity of each agent
can be ignored when one focuses on their interactional structure.
Hence the animals in swarms, the ants foraging, the chemicals interacting in
metabolic systems, the humans in a market, etc., can all be understood as
being comprised of featureless agents and modeled within this
paradigm.

Paradigm II is perhaps as puzzling as the ``unreasonable effectiveness
of mathematics in the natural sciences''. To quote Stephen Wolfram's reaction,
to the realization that simplicity encodes complexity, from  \cite[p. 9]{NKS}:
\begin{quote}
  \it
``And I realized, that I had seen a sign of a quite remarkable and
unexpected phenomenon: that even from very simple programs behavior of
great complexity could emerge.''

``Indeed, even some of the very simplest programs that I
  looked at had behavior that was as complex as anything I had ever
  seen. \\
   It took me more than a decade to come to terms with this
  result, and to realize just how fundamental and far-reaching its
  consequences are.''
\end{quote}

%
%\begin{figure}[tH]
%      \begin{center} \includegraphics[scale=0.38]{complexSys}
%       \caption{\label{fig:c} Simulating a real-world complex system with an agent-based model.}  \end{center}
%\end{figure}
%

In summary, Paradigms I and II can be seen to belong to strategy {\bf
  \textsf B}, as seen in Fig.  \ref{tbl:ovwerlaws}.

\subsection{The Success of Simulating Complex Systems}

\begin{figure}[tH]
       \begin{center} \includegraphics[scale=0.38]{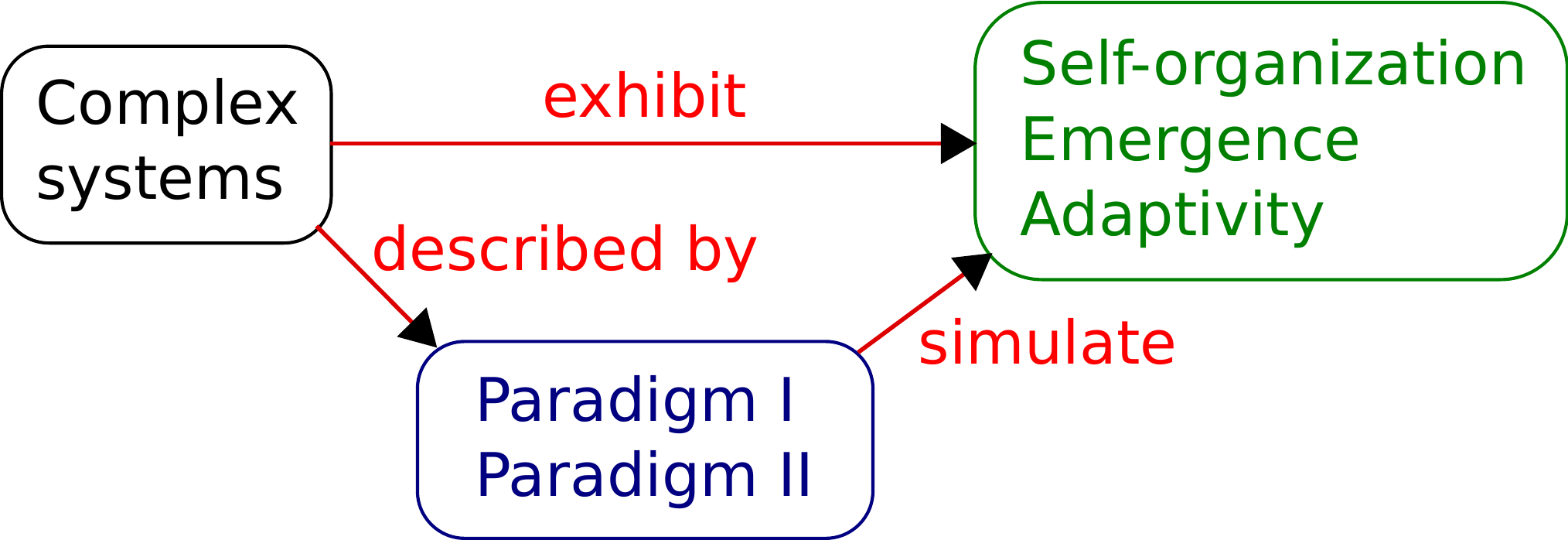}
       \caption{The properties of complex systems and the paradigms describing them.\label{fig:pi-ii}}  \end{center}
\end{figure}

It is remarkable that simple interactions result in complex behavior:
emergence, adaptivity, structure-formation and self-organization. In
essence, complexity does not stem from the number of agents but from
the number of interactions. For instance, there are roughly $30000$
genes in a human vs. about $55000$ genes in a grain of rice.

In Fig. \ref{fig:pi-ii} an illustrated overview of an agent-based
simulation is given: in a computer simulation agents are interacting
according to simple rules and give rise to patterns and behavior seen
in real-world complex systems.

This also highlights the departure from a top-down to a bottom-up
approach to complexity.  As an example, swarming behavior in nature
can be easily modeled by agents obeying three simple and local rules,
reproducing its emergent and adaptive characteristics:
\begin{enumerate}
\item move in the same direction as your neighbors;
\item remain close to your neighbors;
\item avoid collisions with your neighbors.
\end{enumerate}
In addition, biological (temporal-spatial) pattern formation,
population dynamics, pedestrian/traffic dynamics, market dynamics
etc., which where hitherto impossible to tackle with a top-down
approach, are well understood by the bottom-up approach given by the
paradigms of complex systems.

However, to be precise, there is still some mathematical formalism
used in the study of complex systems. For instance, at the macro
level, the so-called Fokker-Planck differential equation gives the
collective evolution of the probability density function of a system
of agents as a function of time. While at the micro level, a single
agent's behavior can be described by the so-called Langevin
differential equation. The two formalism can be mapped into each other
\cite{gardiner1985handbook}. However, as an example, 10000 agents
following Langevin equations in a computer simulation approximate the
macro dynamics of the system more efficiently than an analytical
investigation attempting to solve the equivalent Fokker-Planck
differential equation.

\subsection{An Event-Based Methodology: Intrinsic Time}
\label{sec:it}

By understanding complexity as arising from the interaction of
dynamical systems, it is natural to adopt another paradigm
in their study:
\begin{quote}
\bf
PIII. The passage of time is defined by events, i.e., system interactions.
\end{quote}
In this time ontology\footnote{See also \cite{vanbenthem1991logic}.},
time rests in the absence of events. In contrast to physical time,
only interactions, or events, let a system's clock tick. Hence this
new methodology is called {\it intrinsic time}
\cite{RBO.1983-07-13,UAM.1993-08-16}. Implicit in this approach is
that a system is made up exclusively of interactions, becoming a
dynamic object with a past, present and future. Every interaction with
other systems is a new system. This event-based approach opens the
door to a model that is self-referential, does not rely on
static building blocks and has a dynamic frame of reference.

This approach was a crucial ingredient in the formulation of the new
empirical scaling laws \cite{glattfelder2010patterns}. This is
described in Secs. \ref{sub:csl} and \ref{sub:dc}.

\section{Scaling Laws}
\label{sec:sl}

The empirical analysis of real-world complex systems has revealed
unsuspected regularities, such as scaling laws, which are robust
across many domains \cite{muller:90,mantegna:95,west:97,amaral1998power,albert1999idw,pastorsatorras2001dac,newman2002rgm,garlaschelli2003usr,newman2005power,glattfelder2010patterns}.
This has suggested that universal or at least generic mechanisms are
at work in the structure-formation and evolution of many such systems.
Tools and concepts from statistical physics have been crucial for the
achievement of these findings \cite{DM03,caldarelli2007sfn}.  In
essence, 
\begin{quote} \bf
scaling laws can be seen as laws of nature found for complex
systems.
\end{quote}

A scaling law, or power law, is a simple polynomial functional
relationship
\begin{equation}
f(x) \;\; \propto \;\; x^{-\alpha}.
\end{equation}
Two properties of such laws can easily be shown:
\begin{itemize}
\item a logarithmic mapping yields a linear relationship;
\item scaling the function's argument $x$ preserves the shape of the function $f(x)$, called scale invariance.
\end{itemize}
See for instance \cite{newman2005power,sornette2000critical}.

Scaling-law relations characterize an immense number of natural
processes, prominently in the form of
\begin{enumerate}
\item scaling-law distributions;
\item  scale-free networks;
\item  cumulative relations of stochastic processes.
\end{enumerate}

\subsection{Scaling-Law Distributions}

Scaling-law distributions have been observed in an extraordinary wide
range of natural phenomena: from physics, biology, earth and planetary
sciences, economics and finance, computer science and demography to
the social sciences
\cite{west:97,amaral1998power,albert1999idw,sornette2000critical,pastorsatorras2001dac,bouchaud:00,newman2002rgm,CCRM02,garlaschelli2003usr,gabaix:97,newman2005power,lux2005financial,dimateo:07}.

It is truly amazing, that such diverse topics as
\begin{itemize}
\item the size of earthquakes, moon craters, solar flares, computer files, sand particle, wars and price moves in financial markets;
\item     the number of scientific papers written, citations received by publications, hits on web-pages and species in biological taxa;
\item     the sales of music, books and other commodities;
\item     the population of cities;
\item     the income of people;
\item    the frequency of words used in human languages and of occurrences of personal names;
 \item   the areas burnt in forest fires;
\end{itemize}
are all characterized by scaling-law distributions. First used to
describe the observed income distribution of households by the
economist Vilfredo Pareto in 1897 \cite{pareto1897new}, the recent
advancements in the study of complex systems have helped uncover some
of the possible mechanisms behind this universal law. However, there
is still no conclusive understanding of the origins of scaling law
distributions.  Some insights have been gained from the study of
critical phenomena and phase transitions, stochastic processes,
rich-get-richer mechanisms and so-called self-organized criticality
\cite{bouchaud:00,barndorff-nielsen:01,farmer:04,newman2005power}.

Processes following normal distributions have a characteristic scale
given by the mean of the distribution. In contrast, scaling-law
distributions lack such a preferred scale. Measurements of scaling-law
processes yield values distributed across an enormous dynamic range
(sometimes many orders of magnitude), and for any section one looks
at, the proportion of small to large events is the same. Historically,
the observation of scale-free or self-similar behavior in the changes
of cotton prices was the starting point for Mandelbrot's research
leading to the discovery of fractal geometry
\cite{mandelbrot:63}.

It should be noted, that although scaling laws imply that small
occurrences are extremely common, whereas large instances are quite
rare, these large events occur nevertheless much more frequently
compared to a normal (or Gaussian) probability distribution. For such
distributions, events that deviate from the mean by, e.g., 10 standard
deviations (called ``10-sigma events'') are practically impossible to
observe. For scaling law distributions, extreme events have a small
but very real probability of occurring. This fact is summed up by
saying that the distribution has a ``fat tail'' (in the terminology of
probability theory and statistics, distributions with fat tails are
said to be leptokurtic or to display positive kurtosis) which greatly
impacts the risk assessment. So although most earthquakes, price moves
in financial markets, intensities of solar flares, etc., will be very
small, the possibility that a catastrophic event will happen cannot be
neglected.

\subsection{Scale-Free Networks}
\label{sub:slnetw}

The degree distribution of most complex networks follows a scaling-law
probability distribution $\mathcal P (k) \; \propto \; x^{-\alpha}$,
see also \cite{barabasi:99,albert2002smc,caldarelli2007sfn}.
Scale-free networks are characterized by high robustness against
random failure of nodes, but susceptible to coordinated attacks on the
hubs. Theoretically, they are thought to arise from a dynamical growth
process, called preferential attachment, in which new nodes favor
linking to existing nodes with high degree \cite{barabasi:99}.
Although alternative mechanisms have been proposed \cite{CCRM02}.

\subsection{Cumulative Scaling-Law Relations}
\label{sub:csl}

Next to distributions of random variables, scaling laws also appear in
collections of random variables, called stochastic processes.
Prominent empirical examples are financial time-series, where one
finds empirical scaling laws governing the relationship between
various observed quantities. As an example,
\cite{glattfelder2010patterns} uncovered 18 novel empirical
scaling-law relations, 12 of them being independent of each other.

In finance, where frames of reference and fixed points are hard to
come by and often illusory, these new scaling laws provide a reliable
framework. We believe they can enhance our study of the dynamic
behavior of markets and improve the quality of the inferences and
predictions we make about the behavior of prices. The new laws
represent the foundation of a completely new generation of tools for
studying volatility, measuring risk, and creating better forecasting
and trading models. The new laws also substantially extend the
catalogue of stylized facts and sharply constrain the space of
possible theoretical explanations of the market mechanisms.

See also
\cite{muller:90,mantegna:95,guillaume:97,galluccio:97,dacorogna:01,glattfelder2010patterns}
and Sec. \ref{sub:dc}.

%%%
\section{FX Market as a Complex System}
\label{sec:fx}

The foreign exchange (FX) market can be characterized as a complex
network consisting of interacting agents: corporations, institutional
and retail traders, and brokers trading through market makers, who
themselves form an intricate web of interdependence. With an average
daily turnover of three to four trillion USD \cite{bis:07,bis:10}, and
with price changes nearly every second, the FX market offers a unique
opportunity to analyze the functioning of a highly liquid,
over-the-counter market that is not constrained by specific
exchange-based rules.

What relevance do the insights presented in Sec. \ref{sec:lon} have
for real-world markets, in particular the FX market?

%%%
\subsection{Algorithmic vs. Analytical Approach}
\label{sub:algovsanalyt}

There has been a long history of attempting to understand finance from
an analytical point of view (i.e., within strategy {\bf \textsf A}).
Indeed, the field of mathematical finance has produced a vast body of
mathematical tools characterized by a very high level of abstraction
\cite{OpFuDe,voit2005statistical}.

We believe this is a misleading approach to fundamentally understand
markets and to devise TMs. On the one hand, to make the equations
tractable, often stringent constraints and unrealistic assumptions
have to be imposed. On the other hand, the reality of markets being an
epitome of a complex system is ignored. Hence our shift to strategy
{\bf \textsf B}.
 
We choose to tackle the problem of financial markets in accordance
with Paradigm I, viewing it in terms of interacting agents. Applying
the algorithmic approach given by Paradigm II, entails understanding
the observed market complexity at the macro level arising from the
interaction of heterogeneous agents at the micro level according to
simple rules. The heterogeneity is given for instance by the traders
geographical locations, trading time horizons, risk profiles, dealing
frequencies, trade sizes etc.

In essence, our TMs are agent-based models. An agent is defined by a
position $p_i$, comprised of the set $\{\bar x_i, g_i \}$, where $\bar
x$ is the current average (or entry price) and $g$ is the gearing
(position size). Each position also has a set of event-based and
simple rules. The agent's interaction with each other are constrained
by the price curve.

%%%
\subsection{Event Time: Directional Change Algorithm and Overshoots}
\label{sub:dc}

There is a very concrete application of Paradigm III, the event-based
methodology, to FX time series. As mentioned in Sec. \ref{sec:it}, it
is tightly connected with the existence of cumulative scaling-law
relations.

%%%%
\begin{figure}[Ht]
\centering
\includegraphics[width=0.5\textwidth]{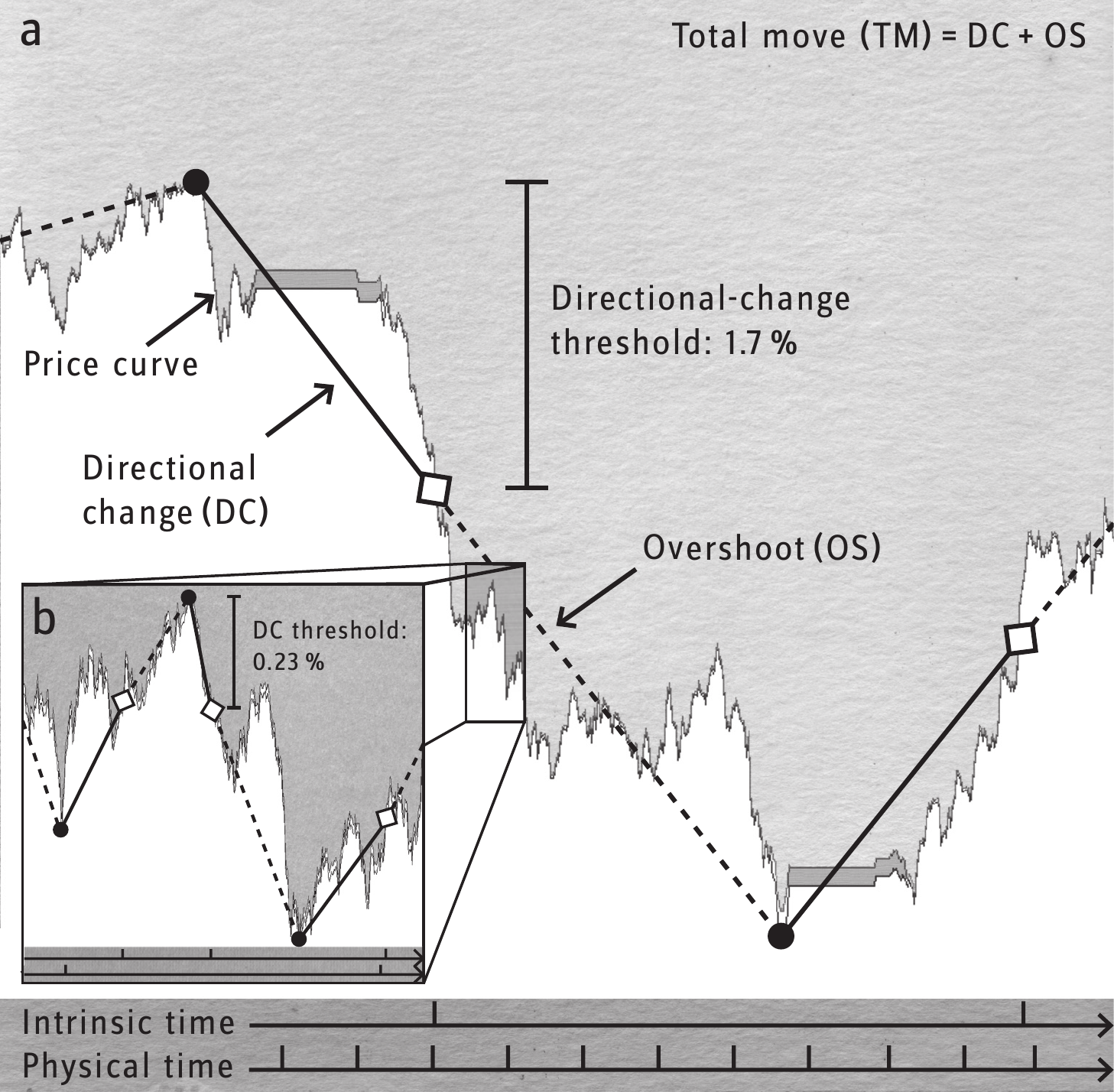}
\caption{ Projection of a (a) two-week, (b) zoomed-in 36 hour price
  sample onto a reduced set of so-called directional-change events
  defined by a threshold (a) $\Delta x_{dc} = 1.7\%$, (b) $\Delta
  x_{dc} = 0.23\%$.  These directional-change events (diamonds) act as
  natural dissection points, decomposing a total-price move between
  two extremal price levels (bullets) into so-called
  directional-change (solid lines) and overshoot (dashed lines)
  sections. Note the increase of spread size during the two weekends
  with no price activity. Time scales depict physical time ticking
  evenly across different price-curve activity regimes, whereas
  intrinsic time triggers only at directional-change events,
  independent of the notion of physical time. Taken from \cite{glattfelder2010patterns}.}
%state.}
\label{fig:dc_to}
\end{figure}
%%%%

In \cite{DMG.1997-01-01} the {\it direction change} algorithm was
introduced. Fig. \ref{fig:dc_to}, taken from
\cite{glattfelder2010patterns}, depicts how the price curve is
dissected into so-called directional-change and overshoot sections.
The dissection algorithm measures occurrences of a price change
$\Delta x_{dc}$ from the last high or low (i.e., extrema), if it is in
an up or down mode, respectively. At each occurrence of a directional
change, the overshoot segment associated with the previous directional
change is determined as the difference between the price level at
which the last directional change occurred and the extrema, i.e., the
high when in up mode or low when in down mode. The high and low price
levels are then reset to the current price and the mode alternates.
\cite{DMG.1997-01-01} presented the directional-change count scaling
law
\begin{equation} \label{eq:dcc}
\mathcal{N} \left( \Delta x_{dc} \right) = \left( \frac{\Delta x_{dc}}{C} 
        \right)^{E},
\end{equation}
where $\mathcal{N} \left( \Delta x_{dc} \right)$ is the number of
directional changes measured for the threshold $\Delta x_{dc}$.

\begin{figure}[tH]
       \begin{center} \includegraphics[scale=0.38]{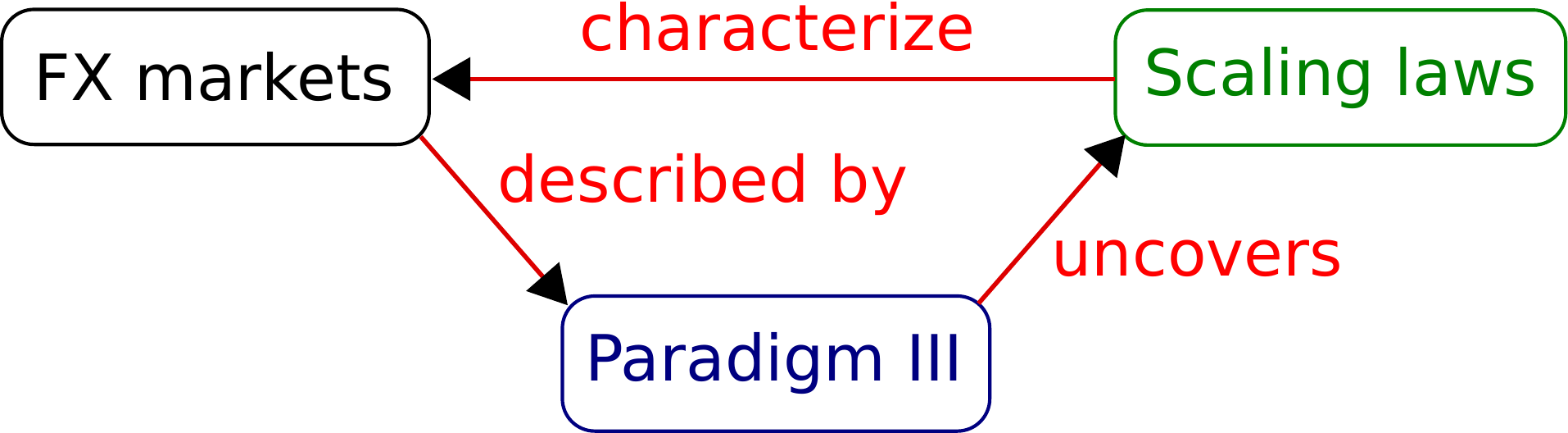}
       \caption{Uncovering scaling laws in the FX market employing Paradigm III.\label{fig:piii}}  \end{center}
\end{figure}

Extending this event-driven paradigm further enables us to observe
new, stable patterns of scaling. In \cite{glattfelder2010patterns}
this event-based approach was crucial for discovering eight of the 12
primary scaling law relations, and four of the six secondary ones.
This establishes the fact, that moving from the empirical time series
to their event-based abstractions provides a unique point of view,
from which patterns can be observed which would otherwise remain
hidden. This is illustrated in Fig. \ref{fig:piii}.

Moreover, the discovery of the overshoot scaling law\footnote{Or more
  precisely: the average overshoot length scaling law, see
  \cite{glattfelder2010patterns}.} was instrumental in extending the
event-based methodology to accommodate a second type of event.  This
scaling law relates the length of the average overshoot segment to the
directional change threshold
\begin{equation} \label{eq:dcc2}
\langle | \Delta x^{os} | \rangle  = \left( \frac{\Delta x_{dc}}{C} 
        \right)^{E}.
\end{equation}
It turns out that the average length is about the same size as the
threshold: $\langle | \Delta x^{os} | \rangle \approx \Delta x_{dc}$.

\begin{figure}[tH]
\centering
\includegraphics[width=0.9\textwidth]{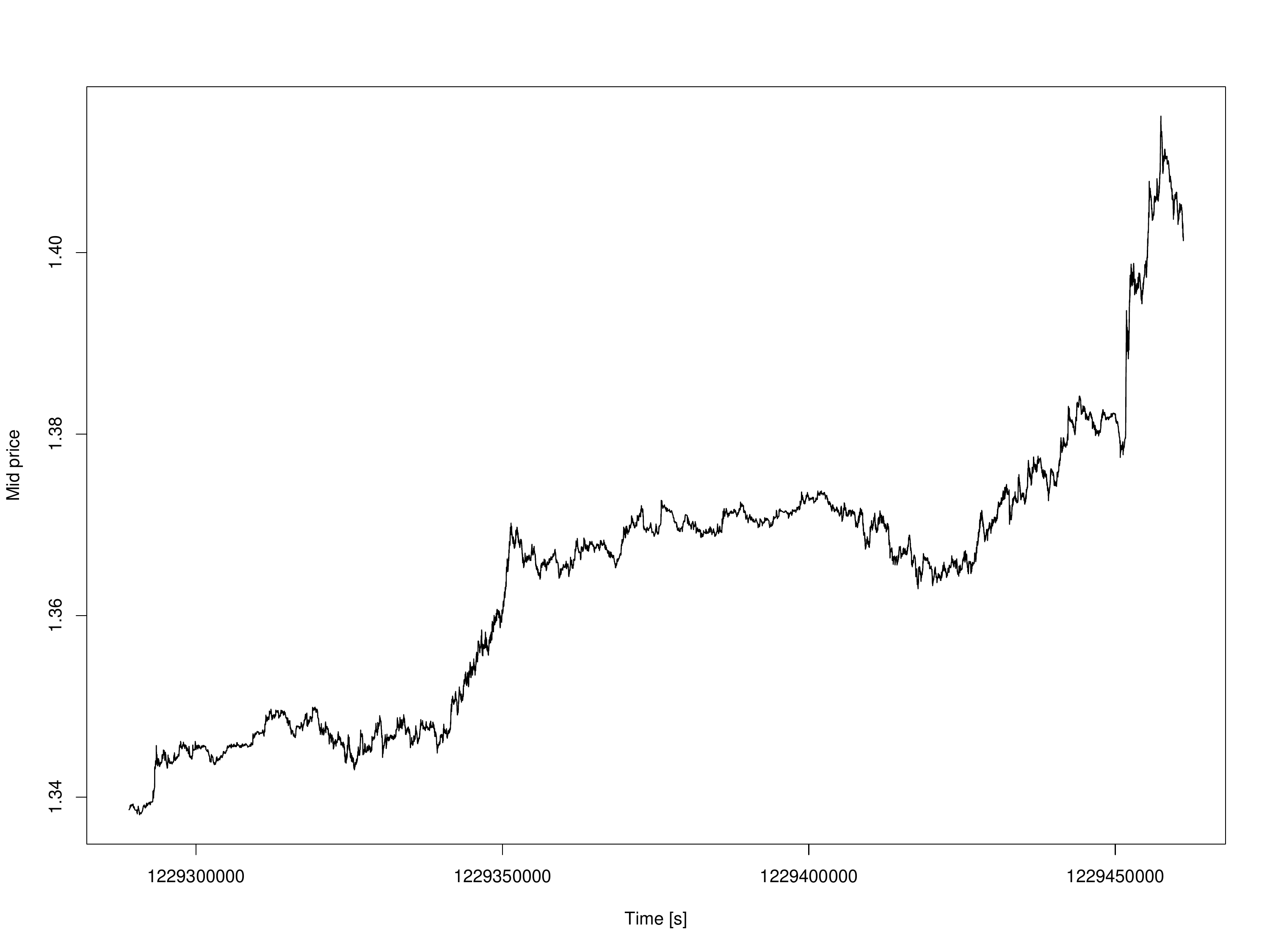}
\includegraphics[width=0.9\textwidth]{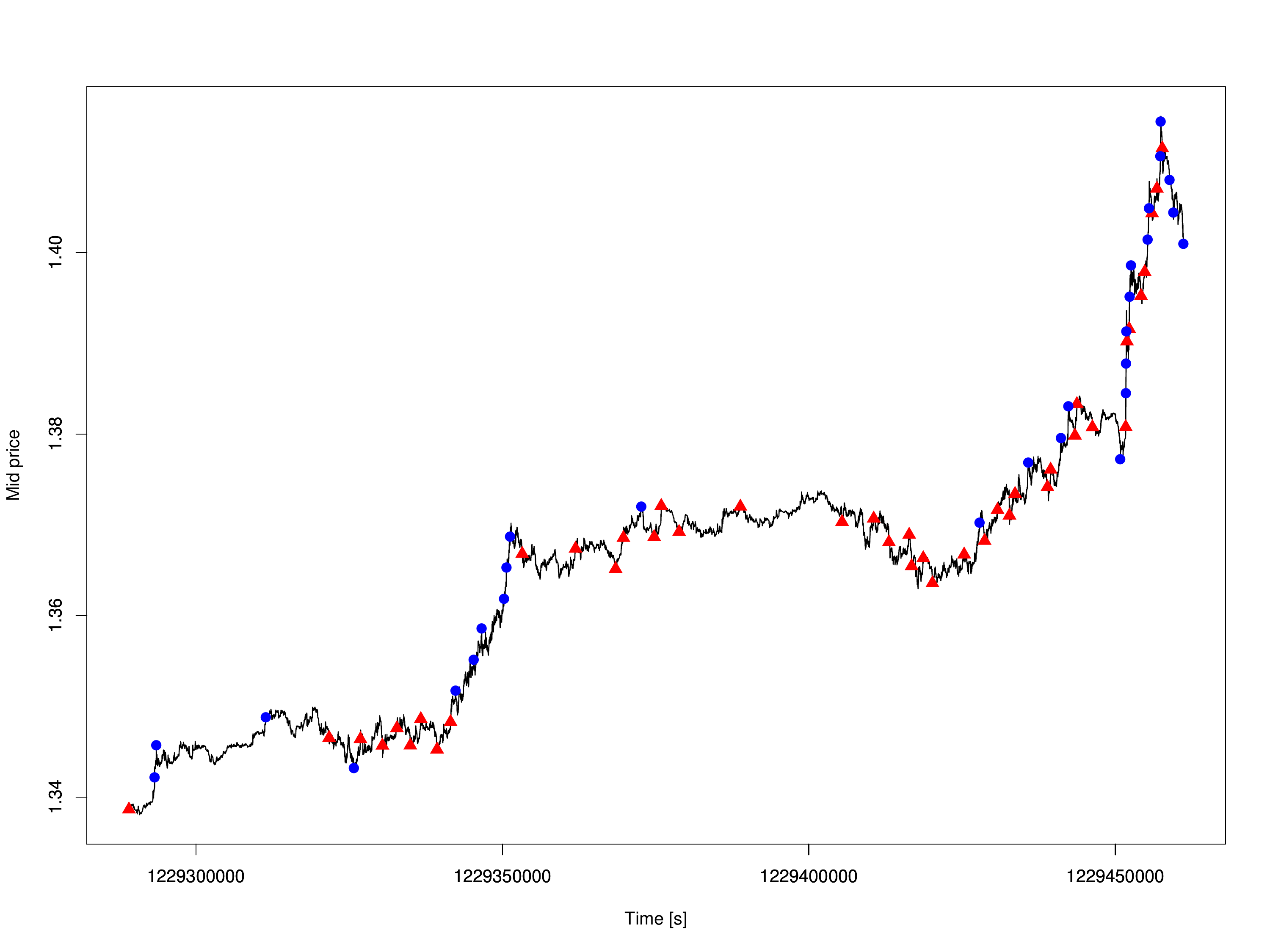}
\caption{({\it Top}) the original EUR\_USD price curve for
  about two days ( 2008-12-14 22:10:56 to 2008-12-16 21:58:20); ({\it bottom}) the price curve is overlaid with
  the directional-change and overshoot events defined by $\Delta
  x_{dc} = 0.25 \%$; the red triangles represent directional-change and the
blue bullets overshoot events.}
\label{fig:cl_plt1}
\end{figure}

\begin{figure}[h!]
\centering
\includegraphics[width=0.9\textwidth]{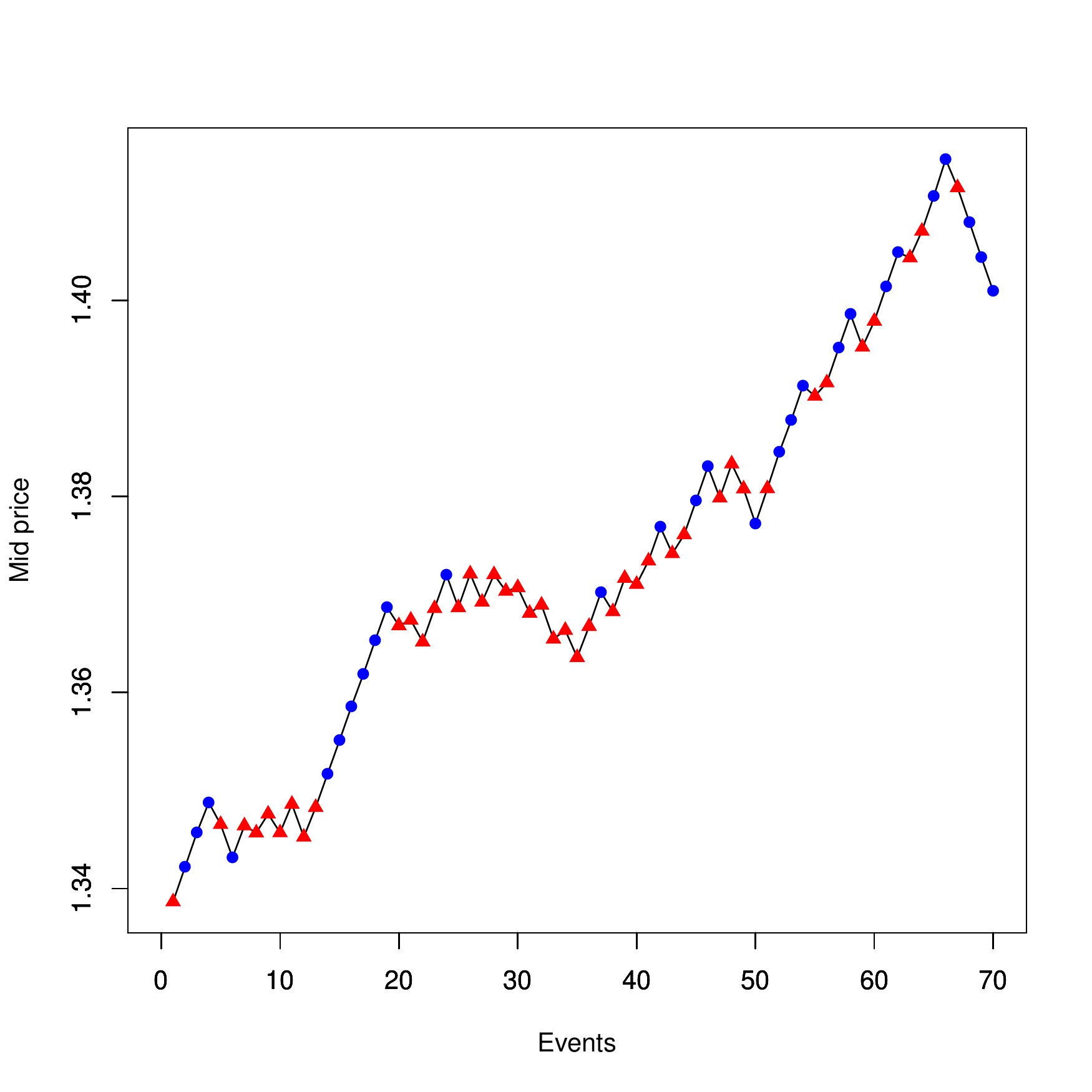}
\caption{The coastline, defined by the threshold of $0.25 \%$, is a
  pure event-based price curve and lacks any reference to physical
  time; its derivation is seen in Fig.  \ref{fig:cl_plt1}; by
  measuring the various coastlines for an array of thresholds, multiple
  levels of event activity are considered.}
\label{fig:cl_plt2}
\end{figure}

This finding motivates the dissection of the price curve into, not
only directional-change events, but also overshoot events, occurring
during the overshoot segments. Crucially, both are defined by the same
threshold $\Delta x_{dc}$. As a result, by applying the direction
change algorithm to empirical price moves, we can reduce the level of
complexity of the real-world time series. In detail, the various fixed
event thresholds of different sizes define focal points, blurring out
irrelevant details of the price evolution. In Fig. \ref{fig:cl_plt1}
an example of an empirical price curve and its associated array of
events for a chosen directional-change threshold is shown. This
effectively unveils the key structures of the market. An example of
this is highlighted in Fig. \ref{fig:cl_plt2}: the {\it coastline},
comprised solely of directional changes and overshoots, with no
constraints coming from physical time. This implies, that the
coastline faithfully maps the activity of the market: during low
volatility, the coastline is shrunk, whereas active market
environments get stretched and all their details are exposed. Hence,
by construction, this procedure is adaptive.

In a nutshell,
\begin{quote} \bf the coastlines associated with different
  directional-change thresholds $\Delta x_{dc}$ are taken as the basis
  for our R\&D effort, coupled with the {\it weltanschauung} coming
  from Strategy {\bf \textsf{B}} and Paradigms I to III.
\end{quote}
\clearpage

%%%%
%%%%

%%%%
%%%%

\clearpage

\pagebreak

%%%%%%%%%

\bibliographystyle{plainnat}

\end{document}